\documentclass[mksc,nonblindrev]{informs3}

\usepackage{graphics}
\usepackage{fourier} 
\usepackage{array}
\usepackage{makecell}
\usepackage{placeins}

\OneAndAHalfSpacedXII

\usepackage{natbib}
 \bibpunct[, ]{(}{)}{,}{a}{}{,}%
 \def\bibfont{\small}%
\TheoremsNumberedThrough     

\EquationsNumberedThrough    

\begin{document}




\TITLE{Data Science for Influencer Marketing : feature processing and quantitative analysis}

\ARTICLEAUTHORS{%
\AUTHOR{Anil Narassiguin, Selina Sargent}
\AFF{Upfluence, \EMAIL{anil.narassiguin@upfluence.com}, \URL{www.upfluence.com}}
} 

\ABSTRACT{%
Influencing a target audience through social media content has become a new focus of interest for marketing leaders. While a large amount of heterogeneous data is produced by influencers on a daily basis, professionals in the influencer marketing (IM) field still rely on simple quantitative metrics such as community size or engagement rate to estimate the value of an influencer. As few research papers have proposed a framework to quantify the performance of an influencer by using quantitative influencer data (number of followers, engagement,...), qualitative information (age, country, city, etc...), natural text (profile and post descriptions) and visual information (images and videos), we decided to explore these variables and quantitatively evaluate their dependencies. By analyzing 713,824 influencers on 5 social media platforms over a period of one year, we identified relationships between value proposition data (engagement, reach, audience), demographics, natural text patterns and visual information. The main goal of this paper is to provide IM professionals with a modern methodology to better understand the value of their influencers and to feed machine learning algorithms for clustering, scoring or recommendation.
}%



\maketitle

%

\section{Introduction}\label{intro}
The last decade has witnessed the proliferation of new paradigms in high-tech fields thank to the emergence of social media platforms. While the big players in IT constantly improve their machine learning algorithms thanks to the huge amount of data generated every day (ref), marketing teams can now rely on social media influencers to better reach their customers. Surprisingly while the two domains have been emerging, almost simultaneously, in the decade of 2010, Influencer Marketing specialists still rely on simple KPIs to find suitable influencer communities for their campaigns. 

IM is a sub-domain of marketing where marketing activities are oriented around an influential person on social media [\cite{brown2008influencer}]. The so-called influencers would promote industry products on their blog posts, Instagram feed, tweets, YouTube videos, etc... Before the information era, the concept of influence was already studied by researchers in sociology, marketing and political science [\cite{katz1966personal}]. 

Since then, with the emergence of the information age and the social networks, many domain-specific studies have been made : case study of IM campaigns [\cite{glucksman2017rise}], relations between IM and brand success [\cite{de2017marketing}], analysis on sponsoring in IM [\cite{ewers2017sponsored}], etc...

Most of these studies have been performed on small subsets of data (hundreds to thousands of influencers) for specific social media and mostly structured data (number of followers, engagement rates, ROIs, etc...). 

Upfluence is an IM company providing services and products for IM professionals. We currently have 2.9 million influencers in our data base and ton better understand how influencers generate enagement, we performed a quantitative analysis on heterogeneous data sources (quantitative data, textual information, visual data). This analysis can also be seen as a methodology for data processing on IM data sets (regular statistics, natural language processing, image recognition, etc...). 

\section{Influencer's value proposition data}\label{basic-features}
In this section, we will review easily processible value information such as numerical data (community size; engagement) and categorical data (geography; gender).

\subsection{Community sizes}
On the main social media (Youtube, Facebook, Instagram, Twitter, Pinterest), the potential value of influencers is traditionally estimated by the size of their community, the number of impressions per post (ie the number of times a post has been displayed to a user),  the interactions between the fan base and the social media content (likes, comments, views) also called engagement and the return on investment. 

The use of certain media has evolved throughout recent years. While Facebook, Twitter and YouTube have been strong references for some time, Instagram is now taking the lead in IM and some platforms such as Pinterest or Twitch are preferred by professionals for specific niche fields (DIY or video games). At Upfluence, the data we gather about different influencers allows us to analyze 572,664 Instagram profiles, 31,526 Facebook profiles, 70,388 Youtube profiles, 24,634 Pinterest profiles and 123,370 Twitter profiles (713,824 unique influencers, as an influencer might have several social networks). 

As presented in table \ref{table:followers}, the average size of community for each social network is highly biased due to extreme values. Most of our influencers have communities expanding thousands to several tens of thousands of followers, which is usually required by IM professionals for marketing campaigns. Recently  business-oriented categories were proposed to better classify influencers according to their community size [\cite{zietek2016influencer}] :

\begin{itemize}
    \item Micro-influencer : less than 15,000 followers
    \item Regular-influencer : between 15,000 and 50,000 followers
    \item Rising-influencer : between 50,000 and 100,000 followers
    \item Mid-influencer : between 100,000 and 500,000 followers
    \item Macro-influencer : between 500,000 and 1,000,000 followers
    \item Mega-influencer : more than 1,000,000 followers
\end{itemize}

Table \ref{table:followers} also shows the distribution of the influencers according to their categories. The business needs are oriented toward Regular and Rising influencers. Indeed large community influencers (Macro and Mega influencers) are usually widely known and therefore have their own connections with quality corporations. Except Pinterest, all the social networks present in the database have a significantly large amount of Regular to Mid influencers which allow us to provide reliable information about the tendencies in IM.

\begin{table}[t]
\centering
\begin{tabular}{cccccc}
\cline{2-6}
\multicolumn{1}{c|}{}         & \multicolumn{1}{c|}{Instagram}   & \multicolumn{1}{c|}{Facebook}    & \multicolumn{1}{c|}{Youtube}    & \multicolumn{1}{c|}{Pinterest}  & \multicolumn{1}{c|}{Twitter}     \\ \hline
\multicolumn{1}{|c|}{Count}   & \multicolumn{1}{c|}{572,664}     & \multicolumn{1}{c|}{31,526}      & \multicolumn{1}{c|}{70,388}     & \multicolumn{1}{c|}{24,634}     & \multicolumn{1}{c|}{123,370}     \\ \hline
\multicolumn{1}{|c|}{Mean}    & \multicolumn{1}{c|}{90,106}      & \multicolumn{1}{c|}{244,013}     & \multicolumn{1}{c|}{304,447}    & \multicolumn{1}{c|}{29,694}     & \multicolumn{1}{c|}{93,275}      \\ \hline
\multicolumn{1}{|c|}{Std}     & \multicolumn{1}{c|}{897,271}     & \multicolumn{1}{c|}{2,215,912}   & \multicolumn{1}{c|}{1,126,288}  & \multicolumn{1}{c|}{199,252}    & \multicolumn{1}{c|}{1,193,945}   \\ \hline
\multicolumn{1}{|c|}{Min}     & \multicolumn{1}{c|}{1,005}       & \multicolumn{1}{c|}{1,000}       & \multicolumn{1}{c|}{1,000}      & \multicolumn{1}{c|}{1,000}      & \multicolumn{1}{c|}{1,000}       \\ \hline
\multicolumn{1}{|c|}{25\%}    & \multicolumn{1}{c|}{14,282}      & \multicolumn{1}{c|}{2,868}       & \multicolumn{1}{c|}{9,031}      & \multicolumn{1}{c|}{1,826}      & \multicolumn{1}{c|}{2,002}       \\ \hline
\multicolumn{1}{|c|}{50\%}    & \multicolumn{1}{c|}{23,427}      & \multicolumn{1}{c|}{8,017}       & \multicolumn{1}{c|}{43,212}     & \multicolumn{1}{c|}{3,841}      & \multicolumn{1}{c|}{4,553}       \\ \hline
\multicolumn{1}{|c|}{75\%}    & \multicolumn{1}{c|}{51,958}      & \multicolumn{1}{c|}{37,611}      & \multicolumn{1}{c|}{195,425}    & \multicolumn{1}{c|}{11,512}     & \multicolumn{1}{c|}{15,200}      \\ \hline
\multicolumn{1}{|c|}{Max}     & \multicolumn{1}{c|}{280,643,232} & \multicolumn{1}{c|}{160,061,728} & \multicolumn{1}{c|}{87,100,920} & \multicolumn{1}{c|}{12,662,073} & \multicolumn{1}{c|}{107,070,300} \\ \hline
                              &                                  &                                  &                                 &                                 &                                  \\ \hline
\multicolumn{1}{|c|}{Micro}   & \multicolumn{1}{c|}{174,120}     & \multicolumn{1}{c|}{19,628}      & \multicolumn{1}{c|}{22,461}     & \multicolumn{1}{c|}{19,559}     & \multicolumn{1}{c|}{92,271}      \\ \hline
\multicolumn{1}{|c|}{Regular} & \multicolumn{1}{c|}{266,073}     & \multicolumn{1}{c|}{4,978}       & \multicolumn{1}{c|}{13,634}     & \multicolumn{1}{c|}{3,069}      & \multicolumn{1}{c|}{16,552}      \\ \hline
\multicolumn{1}{|c|}{Rising}  & \multicolumn{1}{c|}{63,757}      & \multicolumn{1}{c|}{1,862}       & \multicolumn{1}{c|}{7,921}      & \multicolumn{1}{c|}{842}        & \multicolumn{1}{c|}{5,325}       \\ \hline
\multicolumn{1}{|c|}{Mid}     & \multicolumn{1}{c|}{55,473}      & \multicolumn{1}{c|}{3,109}       & \multicolumn{1}{c|}{15,510}     & \multicolumn{1}{c|}{947}        & \multicolumn{1}{c|}{6,292}       \\ \hline
\multicolumn{1}{|c|}{Macro}   & \multicolumn{1}{c|}{6,843}       & \multicolumn{1}{c|}{734}         & \multicolumn{1}{c|}{4,195}      & \multicolumn{1}{c|}{127}        & \multicolumn{1}{c|}{1,204}       \\ \hline
\multicolumn{1}{|c|}{Mega}    & \multicolumn{1}{c|}{6,398}       & \multicolumn{1}{c|}{1,215}       & \multicolumn{1}{c|}{4,623}      & \multicolumn{1}{c|}{90}         & \multicolumn{1}{c|}{1,726}       \\ \hline
\end{tabular}
\caption{Community sizes quartiles per social networks in Upfluence's data extract. Below : Number of influencers per categories}
\label{table:followers}
\end{table}

\subsection{Engagement}

A post's engagement reflects the overall activity and can be defined as the sum of interactions by users on a post. There are various kind of interactions which are dependent on the social platform (Table \ref{table:interactions}). The total sum can be weighed according to the importance given to each interaction. Indeed, a comment given by a follower on an influencer's content implies a priori more engagement than a simple \textit{like}. Even so, in this study we'll suppose that the total engagement is just the simple sum of all the influencer's interactions. 

\begin{table}[t]
\centering
\begin{tabular}{c|c|c|}
\cline{2-3}
                                & Community denomination & Possible interactions                          \\ \hline
\multicolumn{1}{|c|}{Instagram} & Followers              & Likes, comments \\ \hline
\multicolumn{1}{|c|}{Facebook}  & Fans                   & Reactions, comments, shares                     \\ \hline
\multicolumn{1}{|c|}{Youtube}   & Followers              & Likes, dislikes, comments                       \\ \hline
\multicolumn{1}{|c|}{Twitter}   & Followers              & Retweets, favorites, replies                            \\ \hline
\multicolumn{1}{|c|}{Blog}      & Readers                    & Comments, social media shares                                 \\ \hline
\multicolumn{1}{|c|}{Pinterest} & Followers              & Pins, saves, comments, likes                       \\ \hline
\end{tabular}
\caption{Relations between an influencer and his community}
\label{table:interactions}
\end{table}

The total engagement of an influencer is always relative to the size of her/his community. To estimate how an influencer performs relatively to others, the average engagement can be compared to the average time the influencer content has been seen (impression). The engagement rate (ER) is formally a quantity that indicates how much an influencer performs relatively to his visibility on social network [\cite{semiz2017determining}]. It can thus be defined as the following :

\begin{equation*}
    \textsf{engagement\_rate} = \frac{\textsf{\#engagements}}{\textsf{\#impressions}}
\end{equation*}

Even so, the number of impressions for a specific content is hard to estimate, except for the social media companies who have direct access to this data, we usually take the number of followers as an estimate to its potential visibility on social networks :

\begin{equation*}
    \textsf{engagement\_rate} = \frac{\textsf{\#engagements}}{\textsf{\#followers}}
\end{equation*}

\begin{figure}[t]
    \centering
    \includegraphics[width=1\textwidth]{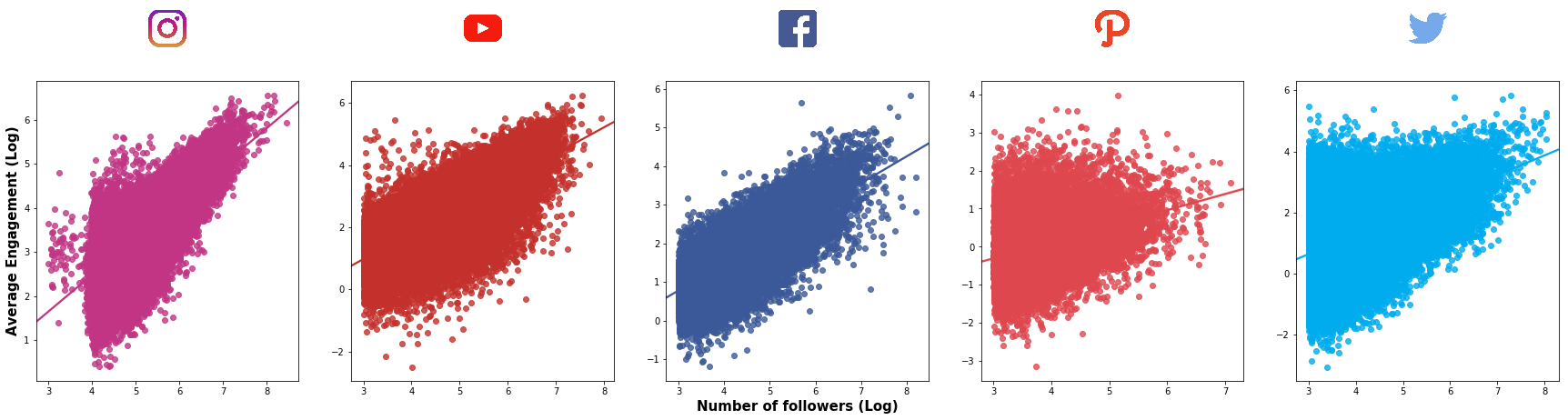}
    \centering
    \caption{Evolution of the average engagement vs the number of followers per social media in logarithmic scale}
    \label{fig:engagement_vs_followers}
\end{figure}

The absolute engagement value has a tendency to disadvantage large community influencers. Indeed, as seen in figure \ref{fig:engagement_vs_followers}, there's a linear evolution between the log of engagement and followers.

Even if this phenomenon is well known by IM specialists, the engagement rate is still used as a reference metric to compare and select potential influencers for marketing campaigns. Practically it may not be bad practice since usually influencers are considered within the same cluster of community size categories (micro, regular, etc...), but this property should orientate the redesign of a new metric to somehow better score influencer performance on specific social media.

Engagement and the engagement rate are actually following log-normal distributions as many human-related behavior metrics (i.e. money distribution across citizens, city population sizes,  length of comments on social media, etc...) [\cite{sobkowicz2013lognormal}]. We verify it empirically by plotting the Q-Q plots of the log of engagement rates per social media (Figure \ref{fig:engagement_density}). 

\begin{figure}[h]
    \centering
    \includegraphics[width=1\textwidth]{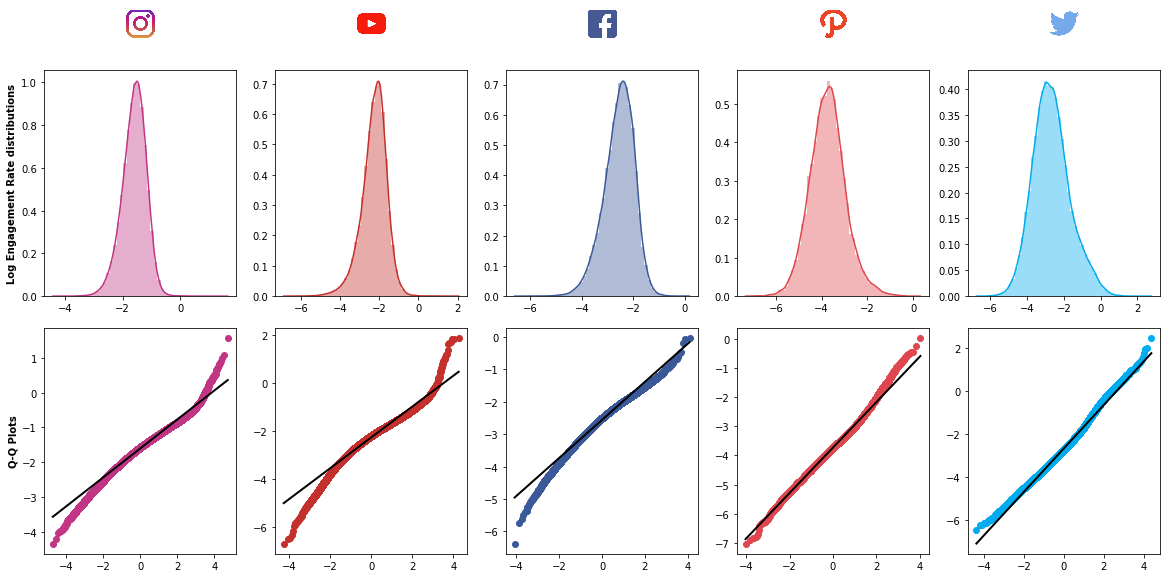}
    \centering
    \caption{Evolution of the average engagement vs the number of followers per social media in logarithmic scale}
    \label{fig:engagement_density}
\end{figure}

This normality allows for the use of a 2/3-sigma rule in order to identify \textit{usual} values for ER as shown in table \ref{tab:ranges}. Depending on the social medium, ER can achieve different ranges of values. Instagram tends to have high engaging influencers while Pinterest tends to have many high community size influencers with low engagement. The later observation has to be deeply studied to compare the potential reach of an influencer on different social media.   

Even though, extreme values for ER don't correspond to specific profiles, we found that by exploring qualitatively some outliers (the 1\% of extreme values), we realised that low ER profiles on YouTube and Facebook usually correspond to brands or news pages (well known organisations, many followers but few real reactions...) while high ER can either be fake profiles or real quality niche influencers. This observations should be considered carefully since we didn't perform any statistical analysis on the categories of outliers' profiles but we hope to do it in a near future.

\begin{table}[t]
\centering
\resizebox{\columnwidth}{!}{
\begin{tabular}{|c|c|c|c|}
\hline
          & Usual range of values (in \%, to be checked with quantile formula)                                                                & Low ER outliers                        & High ER outliers                                                        \\ \hline
Instagram & $3.62 \times 10^{-1}$ - $17.14$                                                 & \thead{Niche communities \\ Caption pictures posts}     & \thead{High quality profiles \\ Scam content \\ Fake engagement \\ Small buzzing influencers} \\ \hline
Facebook  & $1.82 \times 10^{-2}$ - $4.23$                                                  & \thead{News pages \\ Brands pages}                       & Micro influencers                                                               \\ \hline
Youtube   & $2.67 \times 10^{-2}$ - $10.73$                                                 & \thead{Content aggregation websites \\ Education \\ Media} & Buzzing influencers                                                             \\ \hline
Pinterest & $5.13 \times 10^{-4}$ - $6.82 \times 10^{-1}$ &  -  & Popular fields influencers (fitness, DIY, etc...) \\ \hline
Twitter   & $2.03 \times 10^{-5}$ - $2.32 \times 10^{-1}$ & Automatic profiles (reposts, ...) & \thead{Political views \\ Activists \\ Buzzing profiles} \\ \hline
\end{tabular}
\label{tab:ranges}}
\caption {ER usual ranges}
\end{table}

In the rest of the paper, we'll bring to light some interesting correlations between ER and influencer quantitative, qualitative, temporal, textual and visual features. Since the aggregated results are done on large samples of posts and a general selection of influencers, the high differences in engagement rates are usually statistically significant.

This statistical analysis will be an occasion for presenting different data science techniques to extract features from heterogeneous data sources produced by influencers on social media. Indeed, many deep learning techniques allow IM experts to extract meaningful semantics and visual patterns from data that is still ignored by many in the field.

\subsection{Influencer's value proposition data}    \label{fig:er_country}

In this section, we'll see the correlation between the influencers' personal demographic data such as age, sex, place of residence and language spoken, and the influencers' engagement rate. 

\subsubsection{Geographical and language features}

\begin{figure}[t]
    \centering
    \scalebox{.8}{
    \includegraphics[width=1\textwidth]{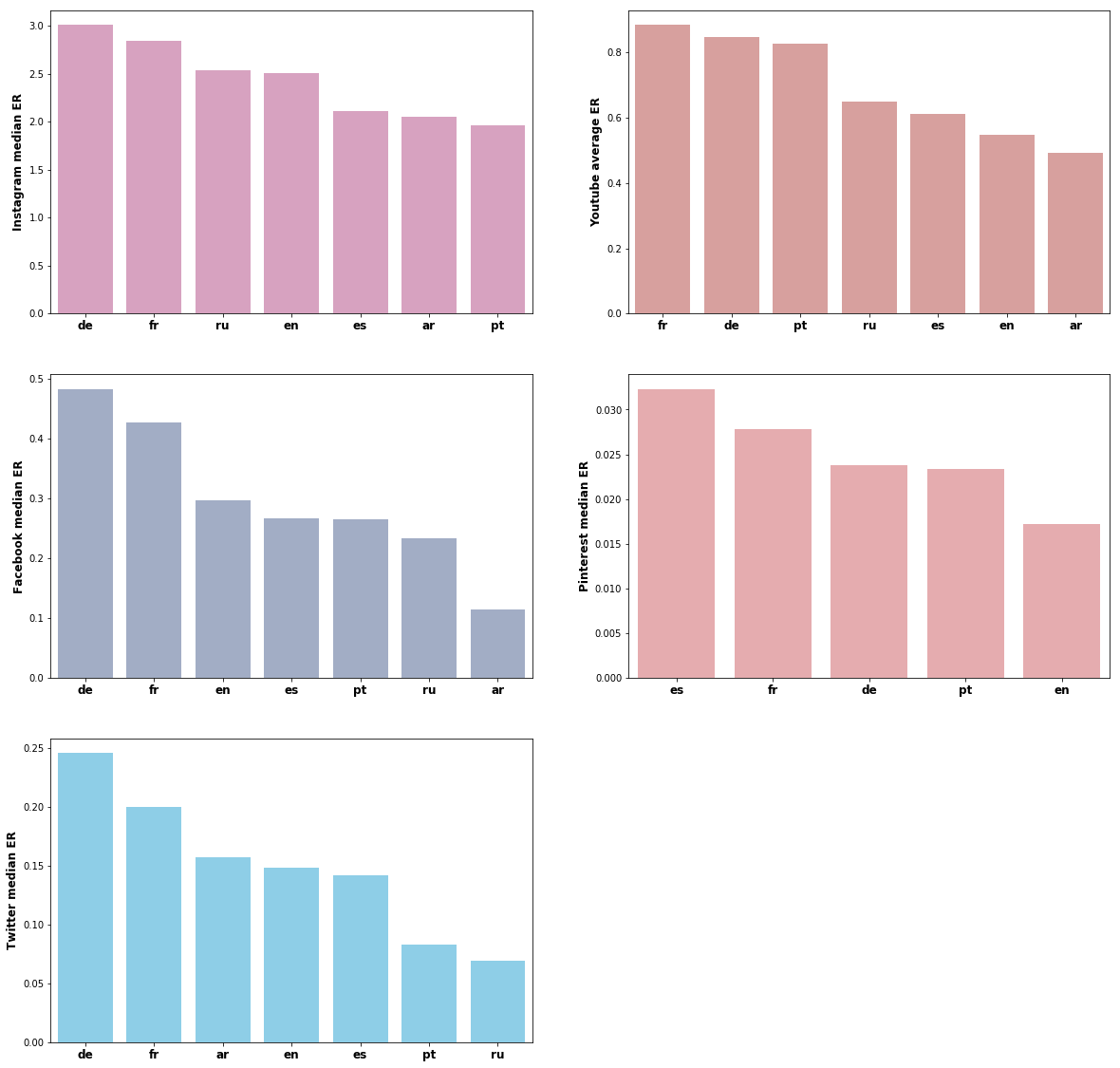}}
    \centering
    \caption{Median engagement rate in percentage per languages and social media}
    \label{fig:er_lang}
\end{figure}

In Figure \ref{fig:er_country} and \ref{fig:er_lang}, these graphs display the rate of engagement with correlation to country and language.  It's worth noting that community size will have,  statistically a significant impact on ER due to country size.

\begin{figure}[t]
    \centering
    \scalebox{.8}{
    \includegraphics[width=1\textwidth]{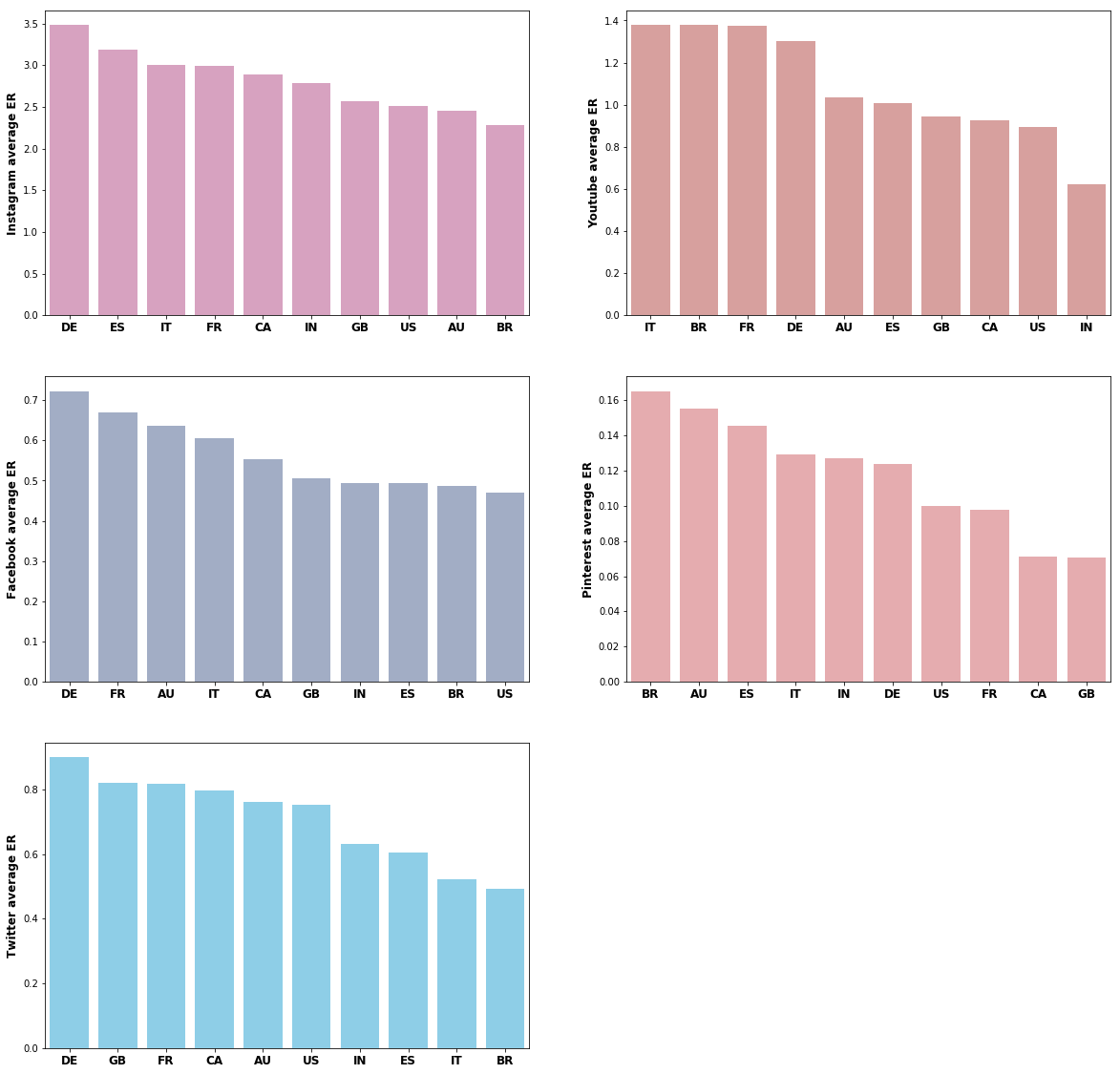}
    }
    \centering
    \caption{Average engagement rate in percentage per countries and social media}
\end{figure}

\subsubsection{Gender}

\begin{figure}[t]
    \centering
    \scalebox{.8}{
    \includegraphics[width=1\textwidth]{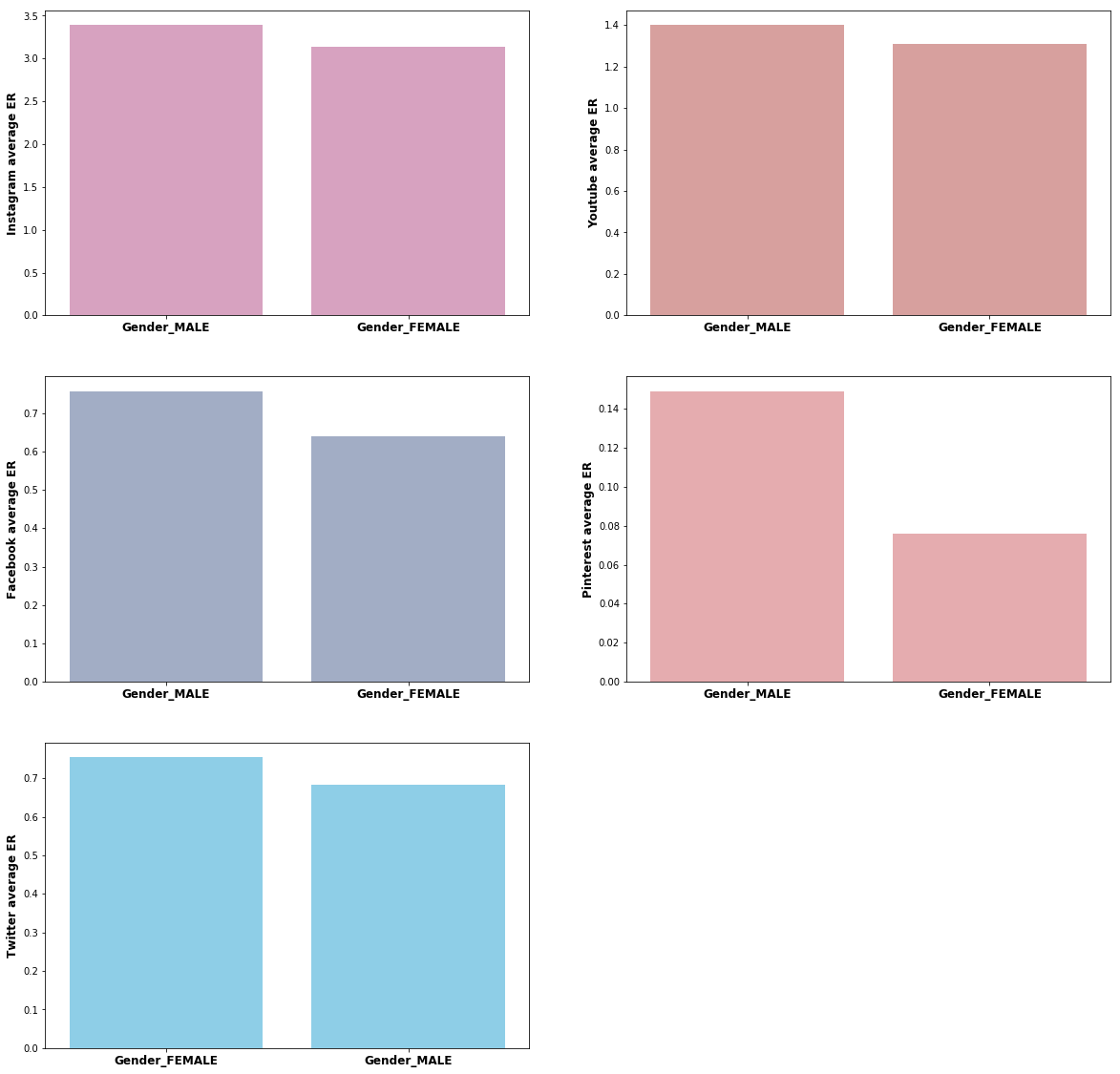}}
    \centering
    \caption{Median engagement rate in percentage per gender and social media}
    \label{fig:gender_er}
\end{figure}

In Figure \ref{fig:gender_er}, for these graphs, we analyzed data collected using image recognition API on social profile pictures. This approach represents an innovative alternative to manual data, however it is supported by gender identification entered manually by the user.  This is enough to get a sufficiently accurate result and allows us to categorise gender identification into two classes.

The nature of the social platform may indicate a gender preference. However, the overall gender split remains similar across all platforms expect Pinterest, where female users are more prevalent (1 man for every 10 women).

\subsection{Temporal behaviors}

\begin{figure}[t]
    \centering
    \includegraphics[width=1\textwidth]{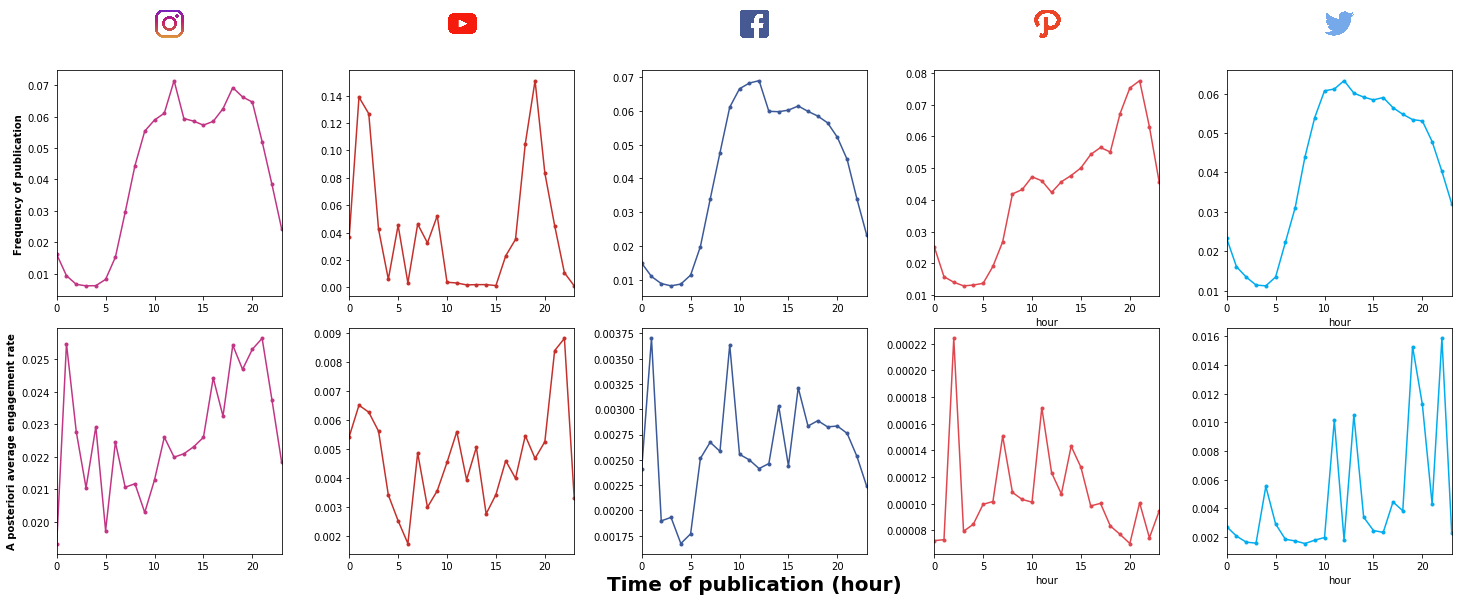}
    \centering
    \caption{Activity and Engagement rate vs Time of the day}
    \label{fig:er_time}
\end{figure}

Figure \ref{fig:er_time} shows that there are trends in our data that suggest the influencer's activity likely correlates to the most active times for their audience across all platforms. Similarities in activity due to social platform algorithms can not be ruled out.

Evidence of ER vs Time show no significant difference, other than on Youtube. It can be reasonably assumed these differences are due to the nature of and use of the platform.

The observed differences can be explained in part by the varying requirements in content creation for Youtube in comparison to the other platforms and in how it is consumed.

The most prevelant use of textual and visual information is available on Instagram, which, outweighs all other platforms when it comes to Influencer preference. Thus, the remaining sections will focus on data collected from Instagram.

\section{Textual information}\label{textual-information}

Textual information has been for decades the first kind of semi structured data available online. Natural language comes with inevitable perceived context which by its nature makes it a difficult medium to infer intent when mined computationally. Recently by using the last neural technologies (word embeddings) for NLP it has been possible to capture strong semantic meaning. Before the emergence of these technologies, tags for the Web and then hashtags for social media, have been a good way for users to specify the content of their posts.  

\subsection{Hashtag vs ER}

\begin{figure}[t]
    \centering
    \scalebox{0.8}{
    \includegraphics[width=1\textwidth]{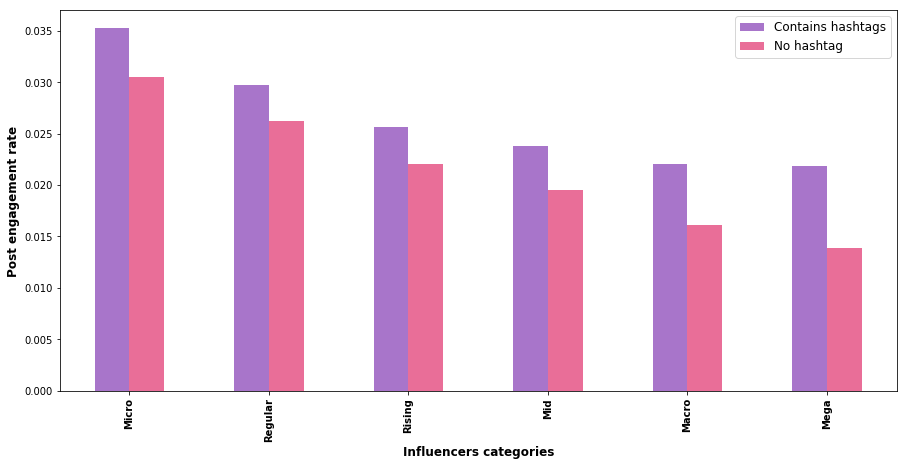}}
    \centering
    \caption{Average engagement rate for posts containing hashtags vs posts containing no hashtag}
    \label{fig:cat_hashtag}
\end{figure}

Primarily used to identify a keyword; topic of interest or to facilitate search, however, increasingly used as a blanket approach in an attempt to increase reach, and can be perceived as a greedy tactic for attention. Therefore posts with zero or demonstrably fewer than excessive use, tend to have higher ER (Figure \ref{fig:cat_hashtag}).

\begin{figure}[t]
    \centering
    \includegraphics[width=0.7\textwidth]{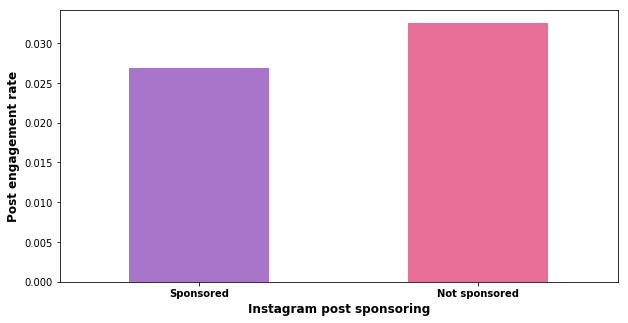}
    \centering
    \caption{Average engagement rate : sponsored posts vs non-sponsored posts}
    \label{fig:er_sponsored}
\end{figure}

In figure \ref{fig:er_sponsored}, we employed a data collected method that included the identification of a predetermined list of specific sponsored hashtags in five languages (French, English, Portuguese, Spanish, German). The correlation showed that ER on posts containing sponsored hashtags was higher in contrast to non-sponsored posts.

Anecdotal (\#natgeo; \#earthpix)  references, as well as mainstream (\#ig\_italy; \#greece) references to travel have significant ER performance.   On the other hand, generic beauty (\#hairstylist; \#nailart) and lifestyle (\#decor; \#chic) related hashtags have a low ER.

\begin{figure}[t]
    \centering
    \includegraphics[width=1\textwidth]{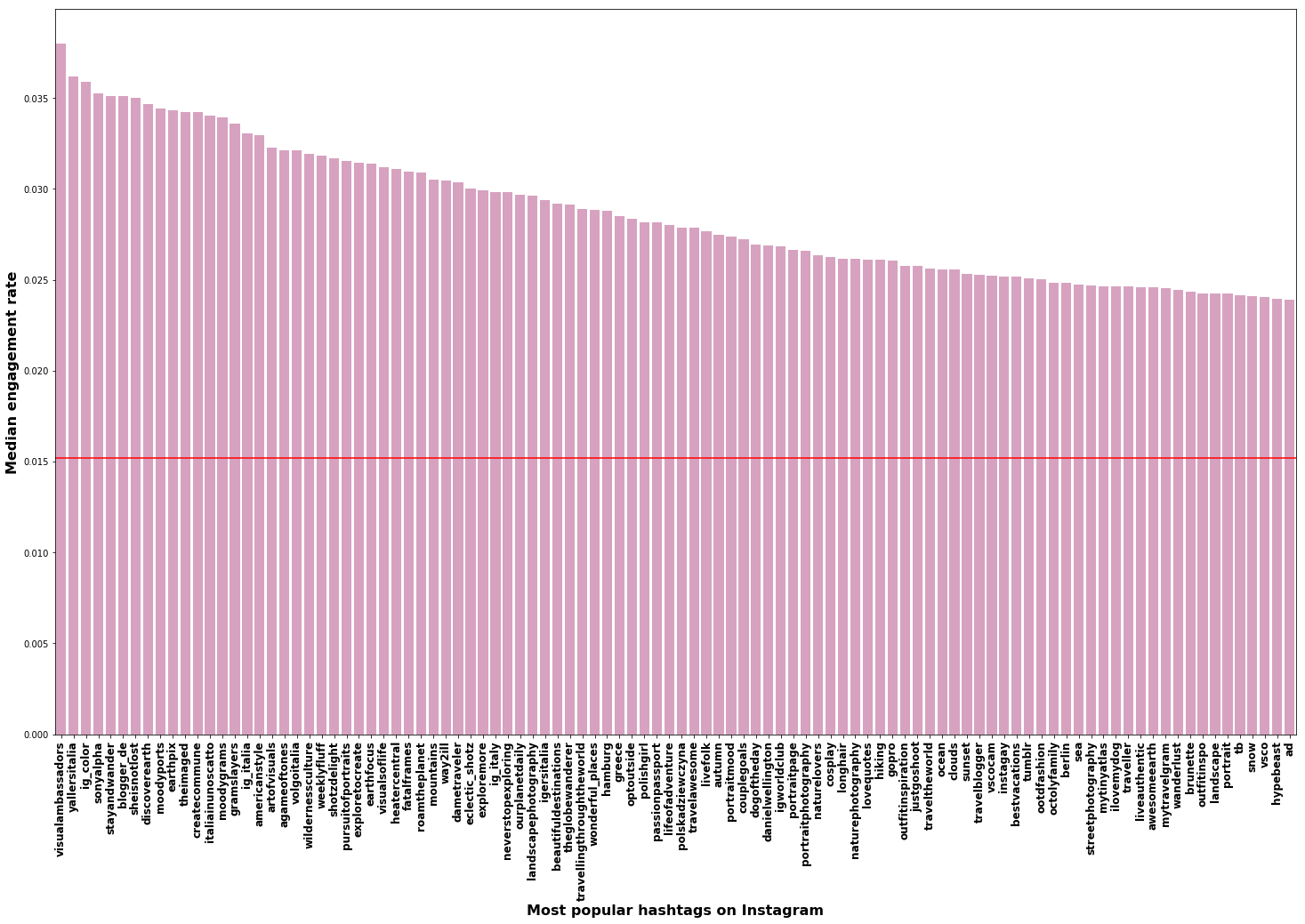}
    \centering
    \caption{Best performing hashtags in terms of ER (red line = overall median ER)}
    \label{fig:engagement_density}
\end{figure}

\begin{figure}[t]
    \centering
    \includegraphics[width=1\textwidth]{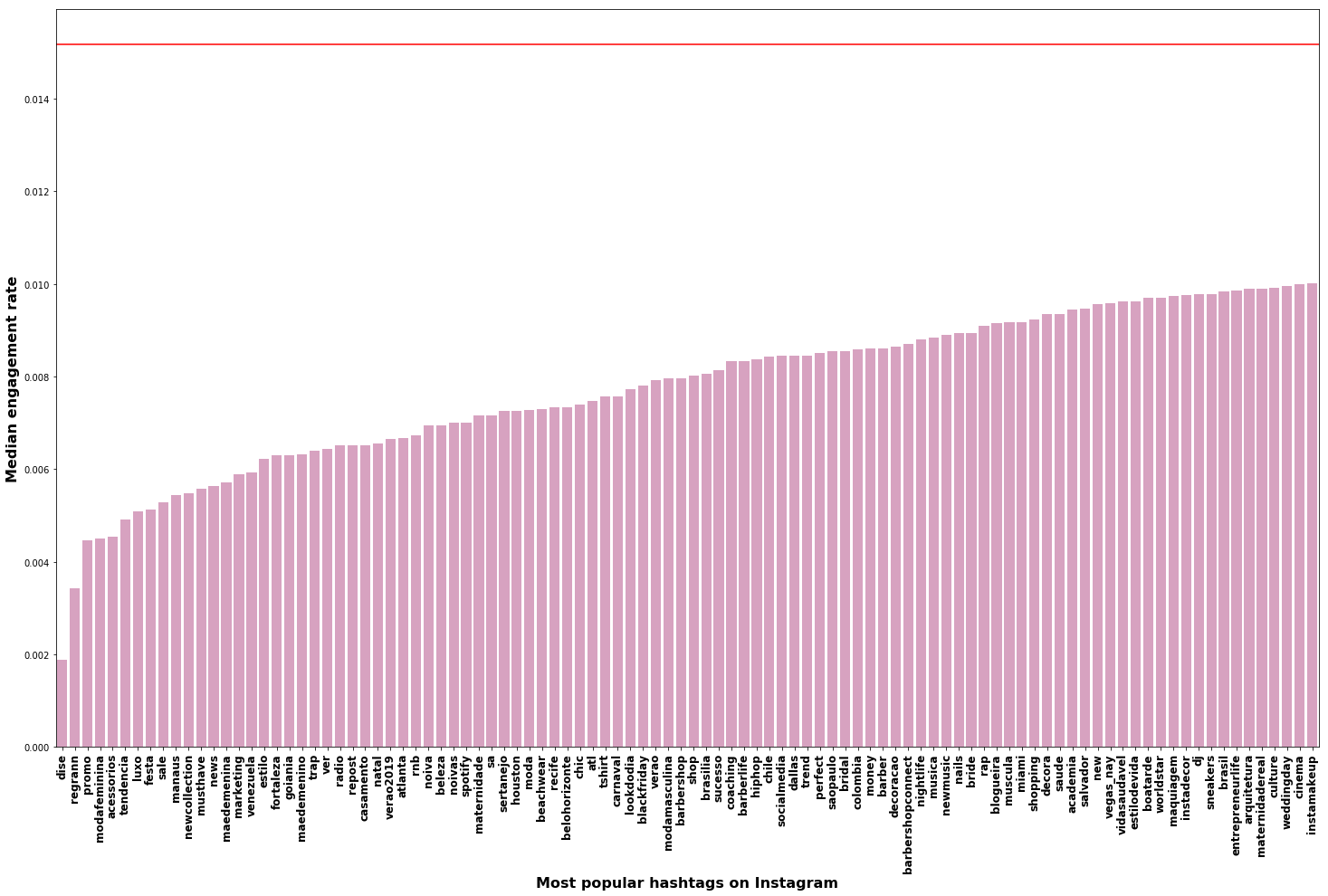}
    \centering
    \caption{Worst performing hashtags in terms of ER (red line = overall median ER)}
    \label{fig:engagement_density}
\end{figure}

\section{Image processing}\label{image-processing}

Since the beginning of the 2010s, image processing techniques have improved consequently thanks to deep learning methods helped by the performance of powerful GPU clusters and the massive amount of unstructured data generated on the Internet. Thanks to image recognition it's possible to detect objects in pictures or video with comparatively high accuracy.

Using a conventional neural network for object detection based on the \textit{YOLO} method [\cite{redmon2016you}] and trained on the Microsoft \textit{COCO} data set [\cite{lin2014microsoft}] it has been possible to automatically tag tens of thousands of images to estimate the average ER for usual object categories (person, dog, cake, etc...).

Our data revealed that profiles with cats, dogs or people had significantly higher ER.

The data isn't significant enough to give a complete conclusion, however, it appears that food related images have lower engagement rates.

\begin{figure}[t]
    \centering
    \includegraphics[width=0.7\textwidth]{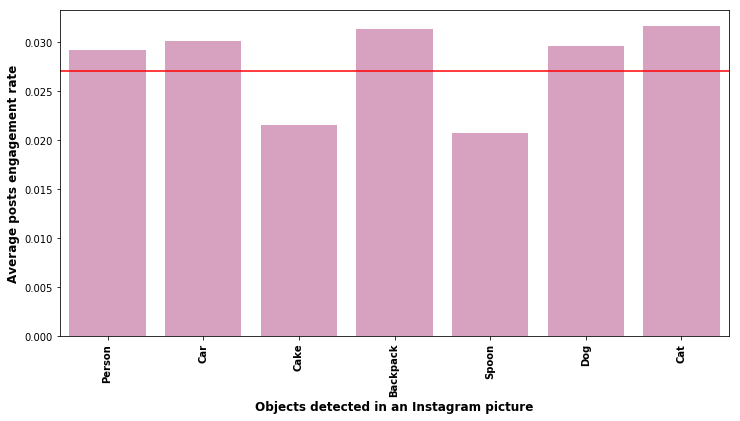}
    \centering
    \caption{Average engagement rate per tags detected in images (red line = overall average ER)}
    \label{fig:engagement_density}
\end{figure}

\clearpage
\section{Conclusion}

Overall, the results support the idea that content is consistent with high engagement, as is demonstrable in certain posts that display a combination of certain visual and textual patterns.   When non-structured data is present, this variation of content can be captured by hashtags, high level deep learning features such as word embedding or convolutional neural networks.  Combined with more readily available influencer information (followers; engagement; country; gender), it gives a more contextual view on the piece of content produced by the influencer.

These findings have significant implications for the understanding of AI and social post data mining.  We presented feature engineering techniques that help IM specialist and data scientist to better represent their data and feed it to machine learning algorithms for recommendations; outlier influencer detection; brand recognition, etc.

However, more research on this topic needs to be undertaken before the association between machine learning techniques and influencer presence on social media is more clearly understood.


%
%
%


\bibliographystyle{informs2014} 
\bibliography{bibliography} 


\end{document}